\documentstyle[11pt]{article}
\def\be{\begin{equation}}
\def\ee{\end{equation}}
\begin{document}
 
\centerline{ \Large{\bf  Einsteins Arbeiten in Bezug auf die moderne Kosmologie}} 

\bigskip

\centerline{  \large{\bf De Sitters L\"osung der Einsteinschen Feldgleichung 
 mit positivem}}

\medskip

\centerline{ \large{\bf  kosmologischen Glied als Geometrie des 
inflation\"aren Weltmodells }} 

\vspace*{0.3cm}

\centerline{Privatdozent Dr. habil. Hans - J\"urgen  Schmidt, Diplommathematiker}
\centerline{http://www.physik.fu-berlin.de/\~{}hjschmi, \ 
e-mail: hjschmi@rz.uni-potsdam.de, Tel. 0331/9771347}
\centerline{Institut f\"ur Mathematik, Universit\"at
Potsdam, Am Neuen Palais 10,  D-14469 Potsdam, Germany}

\bigskip

\noindent
{\bf  Abstrakt:} Die Arbeit [1] von Albert  Einstein von 1918 zu 
Willem De Sitters L\"osung 
[2] der Einsteinschen Feldgleichung wird unter heutigem Gesichtspunkt  
kommentiert. Dazu wird zun\"achst die Geometrie der De Sitterschen Raum-Zeit
beschrieben, sowie ihre Bedeutung f\"ur das inflation\"are Weltmodell erl\"autert.

\medskip

\noindent
{\bf English language title: Einstein's papers in relation to modern cosmology}

\noindent
{\bf Abstract.} We comment on  the paper [1] by Albert Einstein from 1918 
to Willem De Sitter's solution [2] of the Einstein field equation 
from today's point of view. To this end, we start by describing the 
geometry of the De Sitter space-time and present its  importance 
for the inflationary cosmological model. 

\bigskip

\section{Einleitung}

Um die Arbeit  [1] von Albert  Einstein mit dem Titel\footnote{Das Hrn. 
im Titel ist eine Abk\"urzung f\"ur Herrn und nicht f\"ur De 
Sitters Vorname, der lautet Willem.}
 ``Kritisches zu einer von Hrn. De Sitter gegebenen L\"osung 
der Gravitationsgleichungen" angemessen beurteilen zu k\"onnen, mu\ss{} 
man sich klarmachen, da\ss{} im Jahre 1918 die Differentialgeometrie
 der zugrundeliegenden Raum-Zeiten ein noch wenig erforschtes Gebiet war. 
  Man t\"ate Einstein also Unrecht, wenn man mit heutigem Wissen an seine damalige 
Arbeit heranginge, und feststellte, wo \"uberall er   mathematische Fehler begangen hat. 
Vielmehr ist zu beurteilen, welche  Fehler er bei gr\"undlichem 
Literaturstudium h\"atte vermeiden  k\"onnen und welche nicht.  \"Ahnliches 
ist zu den kritisierten Arbeiten [2] De Sitters zu sagen. 

\bigskip

Nachfolgend soll deshalb  zun\"achst in Abschnitt 2 die Geometrie der 
Schwarzschildschen und der De Sitterschen Raum-Zeit 
ausf\"uhrlich  beschrieben werden  (siehe auch [3] bzw. die Lehrb\"ucher [4, 5, 6, 7]),
 sowie in Abschnitt 3 kurz ihre Bedeutung f\"ur das inflation\"are 
Weltmodell  erl\"autert werden (vgl. hierzu wieder  [4-7] sowie 
 die  Arbeiten [8] und [9]). Schlie\ss{}lich sollen in Abschnitt 4 die 
 Arbeiten  [1] und   [10]  von Albert Einstein kurz kommentiert werden. 

\section{Die Geometrie der De Sitterschen Raum-Zeit}

Die De Sittersche Raum-Zeit ist durch folgende Definition eindeutig bestimmt:  sie
ist  die einzige homogene isotrope Raum-Zeit von positiver Kr\"ummung. 
Sie ist die geometrische Grundlage des inflation\"aren Weltmodells, deshalb soll
sie hier detailliert eingef\"uhrt werden. Als Vorbereitung dazu werden zun\"achst 
die Begriffe Koordinatensingularit\"at, echte Singularit\"at, Horizont und 
Schwarzes Loch  gekl\"art. 

\subsection{Koordinatensingularit\"at und echte Singularit\"at}

Generell wird der Begriff Singularit\"at verwendet, um auszudr\"ucken, da\ss{} eine 
Gr\"o\ss{}e ihren zul\"assigen Geltungsbereich verl\"a\ss{}t, in den meisten F\"allen
 geschieht das dadurch, da\ss{} eine Gr\"o\ss{}e, die nur positive 
reelle Werte annehmen darf, gegen Null
oder gegen Unendlich konvergiert. Man unterscheidet  eine 
Koordinatensingularit\"at von einer echten Singularit\"at, je nachdem, ob sich 
diese  dadurch  beseitigen l\"a\ss{}t, da\ss{} man ein anderes Bezugssystem 
verwendet,  oder ob das nicht m\"oglich ist. 

\bigskip

Der einfachste Fall einer Koordinatensingularit\"at ist die
 Euklidische Ebene in Polarkoordinaten $(r, \, \varphi)$ bei $r=0$: 
Der Geltungsbereich dieser Koordinaten ist durch $r > 0$ und
$0 \le \varphi < 2 \pi$ gegeben. Die Einschr\"ankung f\"ur $\varphi$
ist Ausdruck der Tatsache, da\ss{} der Vollwinkel $\varphi = 2 \pi$, 
d.h. $360^0$,
geometrisch nicht vom Winkel $0$ unterschieden wird.\footnote{Die 
topologisch befriedigendere Variante der Polarkoordinaten erlaubt 
allerdings beliebige reelle Werte f\"ur $\varphi$ und nimmt dann eine 
Identifikation aller solcher Winkelwerte vor, deren Differenz ein 
ganzzahliges Vielfaches von $2 \pi$ darstellt. Damit wird verhindert, 
da\ss{} man dem Winkel 0 f\"alschlich eine Sonderrolle 
zukommen l\"a\ss{}t.}

 Die 
Einschr\"ankung f\"ur $r$ ergibt sich daraus,  da\ss{} bei
$r=0$ alle Koordinatenpaare  $(r, \, \varphi)$ demselben Punkt der
Ebene entsprechen, n\"amlich dem Koordinatenursprung, und dies ist 
unzul\"assig, da die Zuordnung zwischen Punkten und Koordinaten 
eineindeutig (d.h. in beiden Richtungen 
eindeutig) sein soll. Wie entscheidet man nun, ob es sich dabei um eine 
echte Singularit\"at handelt? Die Antwort ist bekannt: Der \"Ubergang  zu
kartesischen Koordinaten  $(x, \, y)$, deren Verbindung mit Polarkoordinaten
durch die Formeln
\be
x = r \cdot \cos \varphi  \, , \qquad y = r \cdot \sin \varphi
\ee
gegeben ist, beseitigt diese Mehrdeutigkeit; also ist  $r=0$ nur eine 
Koordinatensingularit\"at: Wie man aus Formel (1) sieht, ist 
bei $r=0$ auch $x=y=0$, und zwar unabh\"angig davon, welchen Wert der
Winkel $\varphi $ dort annimmt. Bei allen anderen Werten ist dagegen 
die Zuordnung zwischen kartesischen und Polarkoordinaten gem\"a\ss{} (1) stets
eineindeutig. 

\bigskip

Quadriert man die Gleichungen aus (1) und addiert sie, ergibt sich der Satz von 
Pythagoras in der elementaren Form
$$
    x^2 + y^2  = r^2      \, ,
$$
die \"aquivalent in der trigonometrischen Form als
$$
 \cos^2 \varphi + \sin^2 \varphi  = 1 
$$
geschrieben werden kann. Soweit der bekannte Schulstoff.

\bigskip

Um gekr\"ummte Raum-Zeiten beschreiben zu k\"onnen, ben\"otigt man den 
Begriff des Riemannschen Raumes und sein Linienelement $ds$. 
Genaueres hierzu l\"a\ss{}t  sich z.B. in den Lehr\-b\"uchern [5, 6, 7]
nachlesen; hier soll es gen\"ugen, wenn wir jetzt die Euklidische Ebene in 
Form eines  Riemannschen Raumes darstellen. Das  Quadrat $ds^2$ des 
Linienelements  l\"a\ss{}t sich dann in kartesischen
Koordinaten als
\be
ds^2 = dx^2 + dy^2  
\ee
schreiben.
Das ist die infinitesimale Form des Satzes von Pythagoras. Vermittels (1) 
transformiert sich  diese Formel (2)  in Polarkoordinaten wie folgt:
\be
ds^2 = dr^2 + r^2 \cdot  d\varphi^2  \,  .
\ee
Das Linienelement bei konstantem Wert $r$ ergibt 
sich\footnote{Andere Herleitung: Der Umfang eines Kreises vom 
Radius $r$ betr\"agt $u =  r \cdot 2 \pi  $, also mu\ss{} f\"ur den Vollwinkel 
$\varphi = 2 \pi$  das Linienelement diesen Wert $u$ ergeben.}
 nach (3) zu
$ds= r \cdot d \varphi$; daraus wird sofort erkennbar, dass bei $r=0$
die \"Anderungen von $\varphi$ keinen Beitrag zu $ds$ leisten, es also dort
eine Koordinatensingularit\"at gibt. 

\bigskip

Bevor wir jetzt den Begriff der echten Singularit\"at kl\"aren k\"onnen, m\"ussen wir 
 kurz erl\"autern, wie sich die Linienelemente (2) und (3) allgemeiner schreiben lassen.
Zun\"achst  werden die Koordinaten mit $x^i$ bezeichnet, wobei $x^0 = t$
die Zeitkoordinate, und die anderen $x^i$ ($i=1,\, 2, \, 3$) die
Raumkoordinaten  darstellen. Dann  wird das Linienelement in die Form 
\be
ds^2 = g_{ij} dx^i dx^j
\ee
gebracht, wobei hier die Einsteinsche Summenkonvention angewendet wird:
\"Uber Indizes, die sowohl in oberer als auch in unterer Position auftreten, wird 
automatisch summiert, ohne da\ss{} das Summenzeichen notiert wird. 

\bigskip

Die Gr\"o\ss{}en $g_{ij}$ sind die Komponenten der Metrik. Sie werden nach 
Allgemeiner Relativit\"atstheorie in einer Doppelrolle verwendet: sowohl zur
Beschreibung der Geometrie der Raum-Zeit  als auch zur Darstellung
des Gravitationsfeldes. Diese Doppelrolle tr\"agt die Bezeichnung: 
``Geometrisierung des Gravitationsfeldes". Konkret hei\ss{}t das zum Beispiel:
Im Satz von Pythagoras steht im Exponenten die Zahl 2; dies gilt 
infinitesimal auch in der Raum-Zeit, deshalb m\"ussen  auf der rechten Seite
von Gleichung (4)  alle $dx^i$ quadratisch auftreten,  und deshalb 
haben die  $g_{ij}$ eben genau zwei Indizes. In der Feldtheorie wird gezeigt, 
da\ss{} diese Anzahl an Indizes genau den Spin des zugeh\"origen Teilchens 
 festlegt. Also: Gem\"a\ss{} Einsteinscher Theorie hat das Graviton\footnote{Das
 Graviton ist  das dem Gravitationsfeld zugeordnete Teilchen, analog ist das
Photon das dem elektromagnetischen Feld zugeordnete Teilchen. Dieses hat
 den Spin 1, da das elektromagnetische Potential $A_i$ nur einen Index tr\"agt.} 
 den Spin 2, weil im Satz von Pythagoras der Exponent 2 auftritt. 

\bigskip

Eine weitere \"Anderung  gegen\"uber der Riemannschen Geometrie erzwingt 
folgender Umstand: Zwar verschmelzen Raum und Zeit in der Raum-Zeit, 
jedoch bleiben raumartige und zeitartige Koordinaten weiterhin unterscheidbar,
 und zwar dadurch, da\ss{} die entsprechenden Anteile 
in $ds^2$ mit unterschiedlichen  
  Vorzeichen eingehen. Man spricht dann von Pseudoriemannscher Geometrie.
 Wie w\"ahlen hier die Variante, in der die raumartigen Anteile ein 
zus\"atzliches Minuszeichen erhalten. 
Die Metrik der speziellen Relativit\"atstheorie lautet dann 
\be
ds^2 = dt^2 - dx^2 - dy^2 - dz^2 \, , 
\ee
wobei die Einheiten so gew\"ahlt sind, da\ss{} die Lichtgeschwindigkeit $c$
den Zahlenwert 1 hat, anderenfalls m\"u\ss{}ten wir in vielen Formeln
noch zus\"atzlich Potenzen von $c$ einf\"ugen. 

\bigskip

Nun k\"onnen wir einige typische Beispiele f\"ur Singularit\"aten angeben: 
Ein Beispiel f\"ur eine echte Singularit\"at ist der Punkt $t=0$ im
expandieren Weltmodell, hier geben wir die einfachste Form eines
r\"aumlich ebenen Friedmannmodells an, welches mit Strahlung 
angef\"ullt ist, also das hei\ss{}e Urknallmodell, auch hot big bang genannt. 
Das Linienelement lautet
\be
ds^2 = dt^2 - t \cdot  \left( dx^2 + dy^2 + dz^2 \right) \, . 
\ee
Der Faktor $t$ vor der Klammer ist Ausdruck der Tatsache, da\ss{}
sich in diesem Modell alle r\"aumlichen Abst\"ande im Laufe der Zeit 
\"andern, und insbesondere bei $t \to 0$ alle L\"angen gegen $0$
konvergieren. Hier l\"a\ss{} sich, anders als bei Gleichung (3),  keine 
Koordinatentransformation finden, die die singul\"are Stelle bei
$t=0$ beseitigen kann. Wie beweist man das? Man kann die Kr\"ummung 
der Raum-Zeit berechnen, und bestimmt dann solche Gr\"o\ss{}en,
die unabh\"angig vom verwendeten Koordinatensystem stets denselben Wert 
annehmen, man nennt sie Invarianten. Es stellt sich heraus, da\ss{}
die Metrik (6) Kr\"ummungsinvarianten besitzt, die bei $t \to 0$ 
divergieren, d.h. gegen
unendlich konvergieren. Die physikalisch orientierte Argumentation ist die
folgende: Die Einsteinsche Feldgleichung lautet
\be
E_{ij} = 8 \pi  \,  G  \,  \cdot \,  T_{ij } \,  , 
\ee
ihre linke Seite ist rein geometrisch definiert, $G$ ist die Gravitationskonstante,
und die Gr\"o\ss{}en $T_{ij }$ messen die physikalischen Eigenschaften der
Materie, z.B. ist  $T_{00 }$ die Energiedichte. In letztere geht nat\"urlich 
gem\"a\ss{}
der Einsteinschen Formel\footnote{Gem\"a\ss{} obiger Vereinbarung $c=1$ 
h\"atten wir hier nat\"urlich einfach $E=m$ schreiben k\"onnen, aber um des optischen 
Wiedererkennungswerts willen soll das $c$ hier einmal stehenbleiben.} 
$$
E = m \cdot c^2
$$
die Ruhmasse des Systems mit ein. In dieser physikalischen 
Blickrichtung hei\ss{}t das:
Der Urknall stellt eine echte Singularit\"at dar, da bei Ann\"aherung  $t$
gegen Null die Energiedichte \"uber alle Grenzen anw\"achst. 

\bigskip

Ganz anders verh\"alt es sich mit der Metrik
\be
ds^2 = dt^2 - t^2 \cdot dx^2 - dy^2 - dz^2\, .
\ee
Scheinbar kann man hier genauso argumentieren:
Bei $t \to 0$ sind in $x$-Richtung alle L\"angen auf Null reduziert.
Genaueres Nachrechnen ergibt allerdings folgendes: Bei $t \to 0$ divergiert 
keine Kr\"ummungsinvariante, und die nach Einsteinscher Feldgleichung
ermittelten Gr\"o\ss{}en $T_{ij}$ verschwinden  sogar alle. 
In der Tat handelt es sich hier um eine Koordinatensingularit\"at, und zwar
ist sie vom selben Charakter wie die oben in Formel (3) behandelte 
Koordinatensingularit\"at der Euklidischen Ebene in 
Polarkoordinaten.\footnote{F\"ur Liebhaber der komplexen Zahlen sei
hier noch folgendes erg\"anzt: Wenn man die raumartigen Koordinaten
mit der imagin\"aren Einheit multipliziert, ergibt sich in allen quadratischen
Ausdr\"ucken ein  zus\"atzlicher Faktor $(-1)$, und man kann dann die
Formeln aus der Elementargeometrie, z.B. Formel (1), anwenden, um die 
Koordinatensingularit\"at zu beseitigen. Bei der anschlie\ss{}enden 
R\"uckg\"angigmachung der Multiplikation mu\ss{}  nat\"urlich der Sinus (sin)
durch den entsprechenden hyperbolischen Sinus (sinh) ersetzt werden etc.} 
In der Tat geht  Metrik (8) durch eine Koordinatentransformation 
in Metrik (5) \"uber, stellt also die materiefreie Minkowskische Raum-Zeit dar;
 genauer gesagt: Metrik (8) repr\"asentiert eine echte  Teilmenge der 
Minkowskisches Raum-Zeit, w\"ahrend Metrik  (5) sie vollst\"andig darstellt. 

\bigskip

Ein anderer Typ von Singularit\"at einer Raum-Zeit kann dann auftreten, wenn
die Koordinaten so gew\"ahlt sind, da\ss{} ein Teilchen bereits nach
endlicher Eigenzeit\footnote{Das ist diejenige Zeit, die eine  von
 diesem Teilchen mitgef\"uhrte Uhr anzeigt.}
 gegen  solche Punkte der Raum-Zeit konvergieren kann, 
deren Koordinaten unendlich  gro\ss{}e Werte annehmen. Als Beispiel 
betrachten wir das expandierende r\"aumlich ebene Weltmodell nach Friedmann 
mit der Metrik 
\be
ds^2 = dt^2 -a^2(t) \cdot \left(  dx^2 + dy^2 + dz^2 \right)  \, .
\ee
F\"ur die Funktion $a(t)$, den kosmischen Skalenfaktor, soll 
gelten: F\"ur alle reellen Zahlen $t$ ist $a(t) > 0$, und $a(t)$ is eine 
monoton wachsende und zweimal stetig differenzierbare\footnote{Diese
Voraussetzung wird ben\"otigt, da die zweiten Ableitungen der Metrik
in die Berechnung der Kr\"ummungsinvarianten eingehen.} 
Funktion. Auf den ersten Blick k\"onnte man annehmen, 
diese Raum-Zeit h\"atte gar keine Singularit\"at. Berechnet man jedoch
die Bahnen von Teilchen, d.h., die Geod\"aten\footnote{Kr\"aftefrei bewegte 
Teilchen bewegen sich l\"angs Geod\"aten, das sind diejenigen Kurven
in der gekr\"ummten Raum-Zeit der Allgemeinen Relativit\"atstheorie, die 
das Analogon der geradlinig gleichf\"ormig bewegten Beobachter  der 
Speziellen Relativit\"atstheorie darstellen.}
 mit Hilfe  der Geod\"atengleichung,
so ergibt sich: Diese Raum-Zeit ist singularit\"atsfrei genau dann, wenn 
\be
\int_{-\infty}^0 \ a(t) dt = \infty
\ee
gilt, siehe z.B. [9]. Anschaulich hei\ss{}t dieses Ergebnis: Wenn die Bedingung 
(10)  nicht erf\"ullt ist, d.h., wenn der 
 kosmische Skalenfaktor zu schnell klein wird, so kann ein Teilchen bereits 
nach endlicher Eigenzeit bis nach $x \to \infty$ gelangen. Wenn dieser Fall auftritt, 
bedarf es weiterer Untersuchungen, ob es sich dabei um eine echte oder um eine
Koordinatensingularit\"at handelt. Wir werden in Abschnitt 2.3. noch einmal auf diese 
Frage zur\"uckkommen.

\subsection{Horizonte und Schwarze L\"ocher}

Beginnen wir mit einem Zitat aus [8]: ``Der am h\"aufigsten diskutierte 
Effekt der Allgemeinen Relativit\"atstheorie ist die Vorhersage der Existenz 
 Schwarzer L\"ocher. Ein Schwarzes Loch ist ein Himmelsobjekt, das 
so schwer ist, da\ss{} selbst das Licht nicht in der Lage ist, die 
gravitative Anziehungskraft zu \"uberwinden. Anders gesagt: Man erkennt es 
daran, da\ss{} ``nichts" zu sehen ist, wenn man hinschaut. Astronomisch 
reale Bilder des Schwarzen Lochs kann es also nicht geben. Man 
kann aber die umgebende Materie sehen, und wenn diese ganz bestimmte
Eigenschaften aufweist, schlie\ss{}t man daraus  auf ein darin befindliches 
Schwarzes Loch."

\bigskip

Um den Begriff eines Schwarzen Lochs mathematisch genauer zu kl\"aren, 
 mu\ss{} man zun\"achst festlegen, was ein Horizont ist. Anschaulich ist der 
Horizont gerade die Grenze  des Teils der Erdoberfl\"ache, den ich 
von meiner Position aus direkt einsehen kann; es handelt sich also um
um eine beobachterabh\"angige Definition, insbesondere brauche ich
 in bestimmten Situationen meine Position nur um wenige Meter zu \"andern, 
um eine merkliche \"Anderung meines Horizonts  ausmachen zu k\"onnen. 
Man stelle sich etwa einen am Nordpol stehenden Beobachter vor, dann besteht 
sein Horizont aus einem n\"ordlichen Breitenkreis, und welcher Breitenkreis das ist,
h\"angt von der Gr\"o\ss{}e des Beobachters ab, im Grenzfall eines unendlich 
gro\ss{}en Beobachters konvergiert dieser Breitenkreis bis an den \"Aquator, 
aber keinesfalls dar\"uberhinaus.  

\bigskip

In der Raum-Zeit $V$ wird analog definiert: Sei $M$ die Menge 
derjenigen Punkte $x$ aus $V$, die die Eigenschaft hat, da\ss{} eine 
kausale  Kurve\footnote{d.h., eine zeitartige oder lichtartige Kurve; eine 
Kurve nennt man zeitartig, wenn sie Bahnkurve eines Teilchens
darstellt, das sich mit Unterlichtgeschwindigkeit bewegt.} 
von $x$ zu einem Punkt der Weltlinie des Beobachters existiert. 
Der Rand von $M$ hei\ss{}t dann der Horizont $W$ von $V$ 
bez\"uglich dieses Beobachters. Bei dieser Definition kann es durchaus
offen bleiben, ob die Punkte, die den Horizont bilden, auch noch zu $M$
geh\"oren sollen oder nicht.  

\bigskip

Anschaulich gesprochen hei\ss{}t das: Wir gehen davon aus, da\ss{}
Information maximal mit Lichtgeschwindigkeit  \"ubermittelt werden kann, 
dann stellt die Menge $M$ die Menge derjenigen Ereignisse dar, von 
denen der Beobachter irgendwann einmal etwas erfahren kann. 

\bigskip

Wenn man jetzt in der
Allgemeinen Relativit\"atstheorie definieren will: ``Der Teil der Raum-Zeit, 
der sich jenseits  des Horizonts befindet, wird Schwarzes Loch genannt.",
so ist damit zun\"achst eine beobachterabh\"angige Definition getroffen worden.
In Formeln sieht das so aus: Das Schwarze Loch ist derjenige Teil der 
Raum-Zeit, der durch 
$$
V \backslash (M \cup W)
$$
gegeben ist. In der hier gew\"ahlten Definition 
wird also der Horizont nicht als Bestandteil des Schwarzen Lochs angesehen.  

\bigskip

 Um die Definition von dieser Beobachterabh\"angigkeit zu befreien, 
 gibt es folgende M\"og\-lich\-keit: Man nimmt an, da\ss{} au\ss{}erhalb 
eines r\"aumlich beschr\"ankten  Gebiets die Raum-Zeit v\"ollig materiefrei ist; 
dann kann man annehmen, da\ss{} die Raum-Zeit asymptotisch flach ist. 
 Es ergibt sich folgendes Ergebnis: Alle hinreichend weit entfernten Beobachter 
haben dann  genau denselben Horizont. Man ordnet dann dieser Menge von 
Beobachtern  den Begriff  ``Beobachter im Unendlichen" zu. Dann 
erh\"alt man die Definition: ``Derjenige Teil der Raum-Zeit, 
der  f\"ur den Beobachter im Unendlichen  jenseits  des Horizonts 
liegt, wird Schwarzes Loch genannt." Damit ist die Beobachterabh\"angigkeit der 
Definition de facto beseitigt.

\bigskip

Es soll  allerdings nicht verschwiegen werden, da\ss{} sich 
f\"ur den Fall allgemeiner Raum-Zeiten,
z.B. einem inhomogenen Weltmodell, welches  auch asymptotisch 
nicht homogen ist, die Beobachterabh\"angigkeit der Definition dessen, was 
als  Schwarzes Loch bezeichnet werden soll, nicht ohne weiteres beseitigen l\"a\ss{}t.
Das hindert jedoch nicht daran, das ``Schwarze Loch im Zentrum unserer Galaxis"
 als zumindest mathematisch wohldefiniert anzusehen: Unser 
Sonnensystem befindet sich n\"amlich so weit au\ss{}erhalb des Zentrums
der Galaxis, da\ss{} man mit guter N\"aherung jeden Beobachter, der sich
innerhalb unseres Sonnensystems befindet, als  Beobachter im Unendlichen
ansehen kann; und unsere Galaxis ist so weit entfernt von anderen
Galaxien, da\ss{} man mit guter N\"aherung  annehmen kann, da\ss{} die 
Raum-Zeit au\ss{}erhalb unserer Galaxis asymptotisch flach ist. 

\bigskip

Die Metrik f\"ur ein Schwarzes Loch l\"a\ss{}t sich nach Trefftz,
hier zitiert aus Einstein [10], Seite 449  wie folgt beschreiben:
\be
ds^2 = \left(  1 + \frac{A}{w} + Bw^2   \right) dt^2 -
\frac{dw^2}{1 + \frac{A}{w} + Bw^2   } - w^2
( d\vartheta^2 + \sin^2 \vartheta d \phi^2) \, .
\ee
Einstein schreibt hierzu: ``Bei negativem $A$ und verschwindendem
$B$ geht dies in die wohlbekannte Schwarzschildsche L\"osung f\"ur 
das Feld eines materiellen Punktes \"uber.
Die Konstante $A$ wird also auch hier negativ gew\"ahlt werden m\"ussen, 
entsprechend der Tatsache, da\ss{} es nur positive gravitierende Massen gibt. 
Die Konstante $B$ entspricht dem $\lambda$-Glied der Gleichung (1a).
Positivem $\lambda$ entspricht negatives $B$ und umgekehrt."
Hierzu sei folgendes erl\"autert: Die genannte Gleichung (1a) ist die
Einsteinsche Gleichung mit kosmologischem Glied $\lambda$, das
heut meist als Gro\ss{}buchstabe Lambda $\Lambda$ geschrieben 
wird. Der Ausdruck
$d\vartheta^2 + \sin^2 \vartheta d \phi^2$ ist das Linienelement 
der Kugeloberfl\"ache, so  da\ss{} sich die Metrik (11) als kugelsymmetrisch 
mit Radialkoordinate $w$ ergibt. 

\bigskip

Mehr zu Einsteins Kommentaren zu Metrik (11) wird in Abschnitt 4 folgen, hier
soll zun\"achst die aktuelle Interpretation der Metrik (11) angef\"ugt werden: 
Heutzutage wird diese  L\"osung meist Schwarzschild-De Sitter-L\"osung 
genannt, bei $B<0$  stellt sie  ein in der De Sitter Raum-Zeit (siehe folgender 
Abschnitt 2.3.) befindliches Schwarzes Loch dar. Bei $B=0$ ist 
Metrik (11) nach \"Anderung auf heute \"ubliche Schreibweise die 
Schwarzschildl\"osung  von 1916
\be
ds^2 = \left(  1 -  \frac{2m}{r}  \right) dt^2 -
\frac{dr^2}{1 - \frac{2m}{r}  } - r^2
( d\vartheta^2 + \sin^2 \vartheta d \phi^2)
\ee
Es mu\ss{} bei dieser Form nat\"urlich noch erg\"anzt werden, da\ss{}
hierbei  die Einheiten so gew\"ahlt werden m\"ussen, da\ss{} die 
Gravitationskonstante $G$ den Wert 1 hat, anderenfalls ist stets 
$m$ durch das Produkt $G \cdot m$ zu ersetzen. \footnote{Ebenso sollte
erg\"anzt werden, da\ss{}  Metrik (12) auch bei negativen Werten $m$ eine
 im Gebiet $r>0$ mathematisch zul\"assige statische kugelsymmetrische 
L\"osung der Einsteinschen Vakuum-Gleichung darstellt,  die jedoch aus 
den genannten physikalischen Gr\"unden hier nicht weiter behandelt werden soll. }

\bigskip

Wir wollen jetzt diese Metrik (12) im Falle $m>0$
 mit der oben angegebene Definition
eines Schwarzen Lochs in Relation setzen. Metrik (12) ist asymptotisch flach,
da sich bei gro\ss{}en Werten $r$ asymptotisch  
$$
ds^2 =  dt^2 - dr^2 - r^2 ( d\vartheta^2 + \sin^2 \vartheta d \phi^2)
$$
ergibt, und das ist genau die flache Minkowskische Raum-Zeit (5) 
in Kugelkoordinaten.  Es ergibt sich, da\ss{}
als Beobachter im Unendlichen jedes Teilchen in Frage kommt, das 
 sich ununterbrochen im Gebiet $r>2m$ aufh\"alt. Die oben definierte Menge 
$M$  derjenigen Punkte $x$, die die Eigenschaft hat, da\ss{} eine 
kausale  Kurve von $x$ zu einem Punkt der Weltlinie des Beobachters 
im Unendlichen  existiert, ergibt sich dann ebenso durch die 
Bedingung $r > 2m$. Der Horizont $W$ ist also der Rand 
des durch $r > 2m$ definierten Gebiets $M$, und das Schwarze Loch ist 
 die Menge derjenigen Punkte, die weder zu $M$ noch zu $W$ geh\"oren. 

\bigskip

Es w\"are  allerdings zu einfach, jetzt zu folgern: $W$ ist also die
durch $r=2m$ definierte Teilmenge der Schwarzschildl\"osung (12),
das Schwarze Loch also das Gebiet $r<2m$. Das hat folgenden Grund:
 Sowohl bei $r=0$ (hier divergiert der Faktor vor $dt^2$) 
als auch bei $r=2m$  (hier divergiert der Faktor vor $dr^2$)
 wird die Metrik (12) singul\"ar, und 
es ist zu kl\"aren, ob es eine echte oder eine Koordinatensingularit\"at ist. 
Bei $r=0$ ist dies ganz einfach zu beantworten: Es gibt eine Kr\"ummmungsinvariante,
 die  im Falle der Metrik (12) den Wert $m^2/r^6$ annimmt, es handelt sich also
um eine echte Singularit\"at. 

\bigskip

Kommen wir nun zu Bereich $r=2m$. Wir vermuten zun\"achst eine 
Koordinatensingularit\"at \"ahnlich wie die bei $r=0$ in Metrik (3), da die 
bekannten Kr\"ummungsinvarianten s\"amtlich regul\"ar sind.  
Wir w\"ahlen jetzt die Eddington-Finkelstein-Koordinaten, hier 
zitiert nach [6]. Dazu wird die Zeitkoordinate $t$ in Metrik (12)
 durch eine neue Zeitkoordinate $v$ ersetzt, die durch die Formel
\be
v = t + r + 2m \ln (r-2m)
\ee
im Bereich $r>2m$ definiert ist. \"Uber die Nebenrechnung
$$
dv = dt + \frac{dr}{1-2m/r}
$$
ergibt sich schlie\ss{}lich die Metrik zu 

\be
ds^2 = \left(1 - \frac{2m}{r} \right) dv^2 - 2 \, dv \, dr - r^2
( d\vartheta^2 + \sin^2 \vartheta d \phi^2) \, . 
\ee
Die Singularit\"at bei $r=0$ ist nat\"urlich geblieben, jedoch ist nun der 
Bereich  $r=2m$ v\"ollig regul\"ar. Der Horizont $W$ des Schwarzen Lochs 
 ist also die durch $r=2m$ definierte Teilmenge der Metrik (14),
wobei die drei anderen Koordinaten alle m\"oglichen Werte durchlaufen. 
 Versucht man jetzt, dieses mittels Formel (13) in Werte f\"ur $t$ 
umzurechnen, stellt sich heraus, (da $\ln 0 = - \infty$ ist), da\ss{}
dies nur bei $t=\infty$ m\"oglich ist. Wir stellen fest: In Metrik (12)
geh\"ort  der Bereich $r=2m$ bei endlichen Werten von $t$ keinesfalls 
zum   Horizont des Schwarzen Lochs.

\subsection{Die De Sitter Raum-Zeit}

Die oben erw\"ahnte Definition der De Sitterschen Raum-Zeit 
als   einzige homogene isotrope Raum-Zeit von positiver Kr\"ummung
soll jetzt genauer erl\"autert werden. Solche homogenen und isotropen R\"aume 
werden in der Literatur oft auch als ``R\"aume konstanter Kr\"ummung" bezeichnet,
und sie sind lokal durch die Angabe einer einzigen Gr\"o\ss{}e, den 
Kr\"ummungsskalar, eindeutig bestimmt. Geometrisch sind sie am einfachsten als 
Teilmenge eines h\"oherdimensionalen flachen Raums darstellbar, und zwar in 
Analogie zur Elementargeometrie: Die  Oberfl\"ache der Einheitskugel ist
der durch die Bedingung  $x^2+y^2+z^2=1$ definierte Teilraum 
des 3-dimensionalen Euklidischen Raumes.  

\bigskip

Wir beschr\"anken uns hier auf 
die vierdimensionale Raum-Zeit, die die L\"osung der Einsteinschen 
Gleichung mit $\Lambda > 0$ darstellt. Die einfachste  Form ist die
als Metrik (11) mit $A=0$ und $B<0$, d.h.\footnote{In dieser
Form ist der Horizont durch $1 + B w^2=0$ definiert.} 
 eine statisch kugelsymmetrische
Form der Metrik. Eine \"ahnlich einfache Form ist die 
als r\"aumlich ebenes Friedmannmodell (9)  mit 
$$
a(t) = e^{H t} \, , \qquad H = \sqrt{\Lambda/3} > 0\, .
$$ 
Die Metrik hat also die Gestalt
\be
ds^2 = dt^2 - e^{2Ht} \left(dx^2 + dy^2 + dz^2    \right).
\ee
Die Bedingung (10) ist nicht erf\"ullt, also enth\"alt diese Metrik eine Singularit\"at. 
Da es ein Raum konstanter Kr\"ummung ist, mu\ss{} es sich hierbei nat\"urlich um 
eine Koordinatensingularit\"at handeln. In der Tat  l\"a\ss{}t
sich die Metrik (15) vermittels einer geeigneten
Koordinatentransformation, Details siehe z.B. in [3], als echten 
Teilraum in ein geschlossenes Friedmannmodell 
einbetten, und dieses Modell hat dann den Skalenfaktor
$$
a(T) = \cosh (HT) = \frac{1}{2} \left( e^{HT} + e^{-HT}  \right). 
$$
Hierbei handelt es sich um eine singularit\"atsfreie Darstellung der
De Sitterschen Raum-Zeit, sie ist zusammenh\"angend, 
einfach zusammenh\"angend, und geod\"atisch vollst\"andig. 

\bigskip

Die \"Ubereinstimmung dieser drei Darstellungen der De Sitterschen Raum-Zeit 
 ist im ersten Moment erstaunlich, da generell eine statisch kugelsymmetrische
Raum-Zeit, ein geschlossenes und ein r\"aumlich ebenes Friedmannmodell ja 
geometrisch  unterscheidbar sind. Es liegt eben an der hohen Symmetrie:
Die Isometriegruppe der De Sitterschen Raum-Zeit ist 10-dimensional, die
der Friedmannmodelle im allgemeinen 6-dimensional,  und je nachdem, welche 
6-dimensionale Untergruppe dieser 10-dimensionalen Gruppe gew\"ahlt wird,
entstehen diese unterschiedlichen Darstellungen. 
\"Ahnlich kann es f\"ur Irritationen sorgen, wenn man einerseits feststellt, 
da\ss{} Metrik (11) zeitunabh\"angig, also statisch ist, w\"ahrend Metrik (15)
ein echt expandierendes Modell darstellt. Dies l\"a\ss{}t sich wie folgt kl\"aren: 
Eine Zeittranslation in (15) hat zur Folge, da\ss{}  $a(t)$  mit einem 
Faktor multipliziert wird, dieser Faktor l\"a\ss{}t sich danach durch eine 
geeignete Multiplikation der Koordinaten $x$, $y$ und $z$ kompensieren.

\bigskip

Das Bild zu diesem Artikel ist aus [3] entnommen und 
 stellt eine vereinfachte Form der De Sitterschen 
Raum-Zeit als geschlossenes Modell dar: $T$ ist die Zeitkoordinate, und
$\Phi $ mit $-\pi \le \Phi \le \pi$  repr\"asentiert eine der drei Winkelkoordinaten. 
Wir starten vom Punkt $T=0$, $\Phi =0$ und fragen, zu welchen Punkten der 
Raum-Zeit man l\"angs einer Geod\"aten gelangen kann. Bei raumartigen Geod\"aten 
(im Bild mit 1 gekennzeichnet) sind alle Geod\"aten geschlossen und treffen sich
 im Antipodenpunkt $T=0$, $\Phi = \pi$, (der nat\"urlich mit 
$T=0$, $\Phi = - \pi$ identifiziert ist).   Lichtartige Geod\"aten 2 und 
zeitartige Geod\"aten 3, die aus  dem Punkt $T=0$, $\Phi =0$ starten,
verbleiben vollst\"andig im Intervall $- \pi/2 < \Phi  < \pi /2$. 
Ergebnis: der Bereich innerhalb der 4 Strich-Punkt-Linien ist der Bereich, zu dem
man von vom Punkt $T=0$, $\Phi =0$ aus vermittels einer Geod\"aten gelangen 
kann. Das ist also keinesfalls die gesamte Raum-Zeit, also gilt, und das ist
sicher ihre eigenartigste  Eigenschaft: 
Die De Sittersche Raum-Zeit ist nicht geod\"atisch zusammenh\"angend, 
obwohl sie  zusammenh\"angend und geod\"atisch vollst\"andig 
ist.\footnote{In der Riemannschen Geometrie
gilt dagegen: in einem zusammenh\"angenden geod\"atisch vollst\"andigen 
Raum lassen sich je zwei Punkte durch einen Geod\"atenabschnitt 
verbinden. Der Grund, weshalb dieses Ergebnis nicht auf die
Pseudoriemannsche Geometrie \"ubertragbar ist, liegt darin,  da\ss{}
beim Beweis f\"ur Riemannsche R\"aume die Kompaktheit der Drehgruppe
ben\"otigt wird, die Lorentzgruppe, das Pseudoriemannsche Analogon 
dazu,  jedoch nicht kompakt ist. }

\bigskip

Beschr\"ankt man sich auf kausale Geod\"aten, so ist das analoge 
Ergebnis  etwas weniger erstaunlich, es gilt: Ein Beobachter, der  f\"ur alle 
Zeiten $T$ im Punkt $\Phi =0$ ruht, kann nicht von allen Ereignissen 
Kenntnis erhalten, es gibt also auch f\"ur ihn einen Horizont. 
Die Lage dieses Horizonts \"andert sich kontinuierlich mit der Lage des 
Beobachters, der Horizont ist also \"uberhaupt nicht beobachterunabh\"angig 
lokalisierbar.

\section{Das inflation\"are Weltmodell}

Das Standardmodell des Universums ist im einfachsten Fall
 durch ein r\"aumlich  ebenes Friedmannmodell (9)  gegeben, das 
bei $t > 0$ zun\"achst 
durch $a(t) = t^{1/2}$, zu sp\"aterer Zeit dann  durch 
$a(t) = t^{2/3}$ spezifiziert ist. Der Exponent $1/2$ gilt
in der Strahlungsphase (dem hei\ss{}en Urknall), der 
Exponent $2/3$ in der heutigen Phase.

\bigskip

Es gibt allerdings eine Reihe von Problemen, die innerhalb dieses
Modells nicht gel\"ost werden k\"onnen. Eines davon ist folgendes: 
Die beobachtetet kosmische Hintergrundstrahlung (also das
inzwischen stark ausgek\"uhlte heute beobachtbare Relikt des Urknalls)
erscheint uns aus allen Richtungen mit ziemlich gleichen Eigenschaften.
 Diese Strahlung stammt 
nach dem Standardmodell allerdings aus Gebieten der Raum-Zeit, die 
zuvor keinerlei Kausalkontakt gehabt haben konnten. 
Es mu\ss{} also  irgendeinen Mechanismus gegeben haben, 
der eine solche Synchronisierung ausl\"ost. 

\bigskip

Es stellt sich heraus, da\ss{} sich dies Problem am einfachsten dadurch
l\"osen l\"a\ss{}t, da\ss{}  man annimmt, da\ss{} zwischen diesen 
beiden Phasen eine endliche Zeit lang eine inflation\"are Phase der kosmischen 
Entwicklung stattgefunden haben mu\ss{}. Und diese wird durch einen
 Skalenfaktor 
$$
a(t) = e^{Ht}
$$
beschrieben, dabei ist $H$ der Hubbleparameter.  Ist $H$ ein positive
Konstante, so ist dies die exakte  De Sittersche Raum-Zeit (15). 
Erlaubt man eine leichte Zeitabh\"angigkeit von $H$, so spricht man 
von einem quasinflation\"aren Modell,  das ebenfalls die Probleme
 des Standardmodells l\"osen kann. 
 Es gibt unterschiedliche Theorien, wie man zu dieser inflation\"aren Phase 
gelangen kann: Z.B. durch Auswirkungen einer h\"oherdimensionalen 
Welt (verschiedene Kaluza-Klein-Modelle), durch Wirkungen eines zus\"atzlichen 
Materiefeldes (Skalarfeld nach Brans und Dicke, Dilatonfeld, 
Higgsfeld u.a.), oder durch die Ber\"ucksichtigung von Quanteneffekten. 

\bigskip

Letztere f\"uhren dann effektiv zu Korrekturtermen h\"oherer Ordnung in der
  Einsteinschen Feldgleichung; siehe z.B. [9] f\"ur Details.  Dabei sind es  die 
Terme vierter Ordnung,
die das Auftreten eine quasi-De Sitter-Phase als transienten Attraktor
 ergeben. Bei diesem Modell werden also weder zus\"atzliche Felder
noch h\"ohere Dimensionen eingef\"uhrt, um die inflation\"are Phase 
zu erzeugen. Da es ein Attraktor ist, ben\"otigt man auch keine speziellen 
Anfangswerte, um die Phase zu haben. Da der Attraktor  transient ist, 
endet  die inflation\"are Phase  auf jeden Fall nach endlicher Zeit, da
die Feldgleichungen  bei gro\ss{}en Werten $t$ einen
\"Ubergang in  die Phase  $a=t^{2/3}$ erzwingen; das 
graceful exit-Problem anderer Inflationsmodelle tritt hier also gar nicht 
erst auf. 

\bigskip

Abschlie\ss{}end zwei Bemerkungen zur Anwendung der De Sitterschen 
Raum-Zeit in der Kosmologie: Erstens: Die Tatsache, da\ss{} sie in der meist 
verwendeten Darstellung (15) gar nicht geod\"atisch vollst\"andig ist, 
wird zwar selten explizit vermerkt, spielt jedoch  kaum eine
Rolle, da inzwischen alle kosmologischen Modelle davon ausgehen, 
da\ss{}  sowohl vor als auch nach der inflation\"aren Phase 
ein  andersgeartetes Expansionsgesetz  gilt, und der Urknall selbst 
  sowieso nicht mit der klassischen Relativit\"atstheorie behandelt werden kann.
Zweitens: Die Lage des Horizonts ist, wie oben gesagt, vom Beobachter 
abh\"angig, anders ist es mit seiner Gr\"o\ss{}e: die ist im homogenen Weltmodell
f\"ur alle ruhenden Beobachter genauso gro\ss{}. Dies wird bei der Theorie
 der Galaxienentstehung angewendet.

\section{Einsteins Arbeit zu de Sitters Weltmodell}

In seiner 
Abhandlung [1] ``Kritisches zu einer von Hrn. De Sitter
gegebenen L\"osung der Gravitationsgleichungen''  schreibt Einstein: 
``Gegen die Zul\"assigkeit dieser L\"osung scheint mir aber ein
schwerwiegendes Argument zu sprechen, das im folgenden dargelegt 
werden soll." Dazu zitieren wir zun\"achst  aus [8]: ``Einstein gibt 
einen `Beweis' daf\"ur an,
 da\ss{} der Horizont eines Schwarzen Lochs von keinem Teilchen 
 \"uberschritten werden kann; beseitigt man den Denkfehler bei Einstein,
 erh\"alt man das korrekte Resultat, da\ss{} ein Teilchen zwar von au\ss{}en nach innen,
nicht aber von innen nach au\ss{}en diesen Horizont queren kann. 
 Das Verst\"andnis dieser Aussage l\"a\ss{}t sich weiter verbessern, wenn man
 sich entschlie\ss{}t, den ``Bereich innerhalb des Horizonts'' 
 in ``Bereich nach dem Horizont''  umzubenennen, eine  
wegen des Relativit\"atsprinzips absolut zul\"assige Vorgehensweise. 
Denn da\ss{} der  ``Bereich nach dem Horizont''  von keinem Teilchen mehr
verlassen werden kann, ist auch ohne Detailkenntnis der 
 Relativit\"atstheorie verstehbar: Der Horizont hei\ss{}t dann 
Gegenwart, und oben genannte Eigenschaft des Horizonts, nur in 
einer Richtung durchschritten werden zu k\"onnen,   dr\"uckt dann  einfach 
die Alltagserfahrung aus, da\ss{} die Vergangenheit nicht mehr zu \"andern ist.

\bigskip

Der genannte Fehler von Einstein wird auch heute noch vielfach wiederholt. 
 Es handelt sich um einen eigenartigen Verdr\"angungsmechanismus: 
Die Formeln werden korrekt aufgeschrieben, es wird explizit gesagt,
 da\ss{} Raum und Zeit hinfort keine Eigenbedeutung mehr haben sondern
 zur Raum-Zeit verschmelzen, und kurz danach bedient man sich unbefangen
 solcher W\"orter wie ``innerhalb''   oder ``danach'', so als ob ihre umgangssprachliche
 Bedeutung auch im Rahmen der Relativit\"atstheorie noch  g\"ultig w\"are.
Bei der genannten Trefftz-schen L\"osung (11) 
 besteht das Problem darin, da\ss{} die mit der Variable $w$ bzw. 
$r$ (von ``radius'') bezeichnete Koordinate am Horizont 
ihren Charakter von raumartig zu zeitartig ver\"andert."

\bigskip

Gehen wir nun etwas detaillierter in die Arbeiten von Einstein: In [1], also
am 7. M\"arz 1918, berechnet er zun\"achst das, was in heutiger Sprechweise 
(siehe oben)  ``Horizont eines im Ursprung ruhenden Teilchens" genannt wird, 
dazu schreibt er: ``Bis zum Beweise\footnote{Anmerkung: Anfang des 20. 
Jahrhunderts war ``Beweise" noch die korrekte Dativform des Wortes 
``Beweis" in Singular.} des Gegenteils ist also anzunehmen, da\ss{} 
die De Sittersche L\"osung in der im Endlichen gelegenen Fl\"ache
$r = \pi R /2$ eine echte Singularit\"at aufweist, d.h. den Feldgleichungen bei 
keiner Wahl der Koordinaten entspricht." An dieser Stelle irrt er zwar, 
jedoch ist sein Vorgehen durchaus nachvollziehbar, und als 
Forschungsansatz auch akzeptabel: er hat sich bem\"uht, die singul\"are
Stelle  durch eine \"Anderung der Koordinaten zu beseitigen, er fand solche
Koordinaten aber nicht. Zudem pa\ss{}te diese L\"osung so gar nicht in 
das Bild, das er sich zum damaligen Zeitpunkt von der Relativit\"atstheorie
gemacht hatte, er dr\"uckt das, also im M\"arz 1918,  wie folgt 
aus: ``Best\"ande die De Sittersche L\"osung
\"uberall zu Recht, so w\"urde damit gezeigt sein, da\ss{} der durch die Einf\"uhrung 
des ``$\lambda$-Gliedes" von mir beabsichtigte Zweck nicht erreicht w\"are."
Erst Jahre sp\"ater sollte er erkennen, da\ss{} dieser Zweck tats\"achlich
 nicht erreicht wurde. 

\bigskip

Kommentieren wir nun abschlie\ss{}end nochmals die Einsteinsche Arbeit [10] 
vom 23. November 1922. Er verwendet  die Bezeichnung
$$
 f_4 = 1 + \frac{A}{w} + Bw^2  
$$
und diskutiert die Nullstellen von $f_4$ in bezug auf Metrik (11). 
Nach unseren Vor\"uberlegungen aus Abschnitt 2 ist  hier zu  folgern: 
Solche Nullstellen ergeben  einen Horizont und keine echte 
Singularit\"at. Einstein schreibt jedoch: ``Nach (1) ist $\sqrt{f_4}$
 die Ganggeschwindigkeit einer Einheitsuhr, welche an jenem Orte ruhend
angeordnet wird. Das Verschwinden von $f_4$ bedeutet also eine wahre 
Singularit\"at des Feldes."  Gleichwohl ist diese \"Au\ss{}erung Einsteins 
 nicht ausdr\"ucklich falsch, vielmehr ist diese Diskrepanz ein Ausdruck 
der Tatsache, da\ss{} heutzutage f\"ur die Regularit\"at einer Raum-Zeit nicht 
mehr gefordert wird, da\ss{} ruhende Einheitsuhren als 
Zeitvergleichsapparatur m\"oglich sein sollten, wie dies Einstein offensichtlich 
noch gefordert hat. 

\section*{Literatur}

\noindent 
[1] Einstein, A.: Kritisches zu einer von Hrn. De Sitter 
gegebenen L\"osung der Gravitationsgleichungen, Sitzungsberichte der 
K\"oniglich Preussischen Akademie der Wissenschaften, Phys.-Math. Klasse,
1918, Seite 270-272. 

\medskip 

\noindent 
[2] De Sitter, W.: Over de Kromming der ruimte, Koninklijke Akademie
van Wetenschappen te Amsterdam {\bf 26} (1917) 222-236. Die englischsprachige
Variante des original niederl\"andischen Texts erschien  als: On the curvature of 
space, Proc. Amsterdam {\bf 20} (1918) 229.  De Sitter, W.: On Einstein's theory 
of gravitation, and its astronomical consequences, Monthly Notices 
Royal Astron. Soc. {\bf 76} (1916) 699. In denselben Zeitschriften erschienen 
auch eine ganze Reihe von weiteren Arbeiten de Sitters zu \"ahnlichen Themen.   

\medskip 

\noindent 
[3]   Schmidt, H.-J.: On the de Sitter space-time - the geometric
foundation of inflationary cosmology, Fortschr. Phys. {\bf  41} 
(1993) 179-199.

\medskip 

\noindent 
[4]  Schutz, B.: Gravity from the ground up, Cambridge University Press 2003.

\medskip 

\noindent 
[5] Stephani, H.: General Relativity, Cambridge University Press 1982.
Die deutschsprachige Originalausgabe erschien als: 
Allgemeine Relativit\"atstheorie, Verlag der Wissenschaften Berlin 1977. 
 
\medskip 

\noindent 
[6] Hawking, S. W., Ellis, G. F. R.: The large scale structure of space-time, 
 Cambridge University Press 1973.

\medskip 

\noindent 
[7] Landau, L. D., Lifschitz, E. M.: Klassische Feldtheorie, Akademieverlag Berlin 
1962, \"Ubersetzung aus dem Russischen, Originaltitel: Teoria polja, Verlag Nauka, 
Moskau 1958.  

\medskip 

\noindent 
[8] Schmidt, H.-J.: Zur Beweiskraft von Bildern in Mathematik und Astrophysik, 
Beitrag zur Monographie: ``Im Zwischenreich der Bilder", Hrg.: 
Jacobi, R., Marx, B., Strohmaier-Wiederanders, G., zugleich Band {\bf 35}
der Reihe ``Erkenntnis und Glaube", Evangelische Verlagsanstalt Leipzig 2004, 
Seite 267-276. 

\medskip 

\noindent 
[9] Schmidt, H.-J.: Stability and Hamiltonian formulation of higher
derivative theories, Phys. Rev.  {\bf  D49} (1994) 6354-6366; 
 Erratum Phys. Rev.  {\bf  D54} (1996) 7906; diese Arbeit ist als  
Preprint gr-qc/9404038 in www.arxiv.org einsehbar. 

\medskip 

\noindent 
[10] Einstein, A.: Bemerkung zu der Abhandlung  von E. Trefftz:
 ``Das statische Gravitationsfeld zweier Massenpunkte in der 
Einsteinschen Theorie",  Sitzungsberichte der 
 Preussischen Akademie der Wissenschaften, Phys.-Math. Klasse,
1922, Seite 448-449. 

\bigskip

\noindent
\copyright (2005) Hans-J\"urgen Schmidt

\end{document}